\documentclass[%
  journal=jctcce
]{achemso}

\usepackage[utf8]{inputenc}
\usepackage[T1]{fontenc}
\usepackage{xcolor}
\definecolor{blue}{RGB}{0,0,150}
\usepackage[%
  colorlinks=true,
  allcolors=blue
]{hyperref}
\usepackage{url}
\usepackage{amsfonts}
\usepackage{amssymb}
\usepackage{amsmath}
\usepackage{mathtools}
\usepackage[ruled,vlined]{algorithm2e}
\usepackage{enumitem}
\usepackage{lmodern}

\DeclarePairedDelimiter\paren{(}{)}              
\DeclarePairedDelimiter\ang{\langle}{\rangle}    
\DeclarePairedDelimiter\bkt{[}{]}                
\DeclarePairedDelimiter\set{\{}{\}}              
\DeclarePairedDelimiter\enorm{\lVert}{\rVert^2}  

\DeclareMathOperator*{\argmin}{arg\,min}

\newcommand{\eq}[1]{Eq.~(\ref{#1})}
\newcommand{\dd}[1]{\operatorname{d#1}}
\newcommand{\bz}{\mathbf{z}}
\newcommand{\bx}{\mathbf{x}}

\newcommand{\dz}{\dd{\mathbf{z}}}
\newcommand{\dx}{\dd{\mathbf{x}}}

\newcommand{\e}{\operatorname{e}}
\newcommand{\kT}{k_{\mathrm{B}}T}

\title{%
  Reweighted Manifold Learning of Collective Variables from Enhanced Sampling Simulations}

\author{Jakub Rydzewski}
\email{jr@fizyka.umk.pl}
\affiliation{%
  Institute of Physics,
  Faculty of Physics, Astronomy and Informatics,
  Nicolaus Copernicus University,
  Grudziadzka 5, 87-100 Toru\'n, Poland
}

\author{Ming Chen}
\affiliation{%
  Department of Chemistry,
  Purdue University,
  West Lafayette, Indiana 47907, USA
}

\author{Tushar K. Ghosh}
\affiliation{%
  Department of Chemistry,
  Purdue University,
  West Lafayette, Indiana 47907, USA
}

\author{Omar Valsson}
\affiliation{%
  Department of Chemistry,
  University of North Texas,
  Denton, Texas 76201, USA
}

\begin{document}

\begin{tocentry}
  \includegraphics{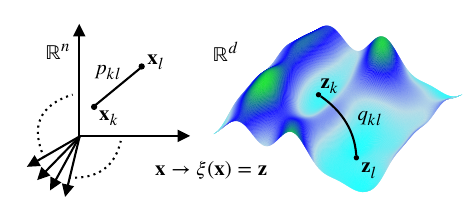}
\end{tocentry}

\newpage

\begin{abstract}
Enhanced sampling methods are indispensable in computational chemistry and physics, where atomistic simulations cannot exhaustively sample the high-dimensional configuration space of dynamical systems due to the sampling problem. A class of such enhanced sampling methods works by identifying a few slow degrees of freedom, termed collective variables (CVs), and enhancing the sampling along these CVs. Selecting CVs to analyze and drive the sampling is not trivial and often relies on chemical intuition. Despite routinely circumventing this issue using manifold learning to estimate CVs directly from standard simulations, such methods cannot provide mappings to a low-dimensional manifold from enhanced sampling simulations as the geometry and density of the learned manifold are biased. Here, we address this crucial issue and provide a general reweighting framework based on anisotropic diffusion maps for manifold learning that takes into account that the learning data set is sampled from a biased probability distribution. We consider manifold learning methods based on constructing a Markov chain describing transition probabilities between high-dimensional samples. We show that our framework reverts the biasing effect yielding CVs that correctly describe the equilibrium density. This advancement enables the construction of low-dimensional CVs using manifold learning directly from data generated by enhanced sampling simulations. We call our framework reweighted manifold learning. We show that it can be used in many manifold learning techniques on data from both standard and enhanced sampling simulations.
\end{abstract}

\maketitle

\newpage

\section{Introduction}
Among the main challenges in atomistic simulations of chemical systems is the significant temporal disparity between the timescales explored in standard atomistic simulations and the long timescales observed in experiments. Atomistic simulations can only reach timescales of up to milliseconds and thus cannot exhaustively sample the high-dimensional phase space, leading to the so-called sampling problem that has both theoretical and computational consequences for dynamical systems. The reason for the sampling problem is that these systems are characterized by many metastable states (i.e., high-probability regions) separated by energy barriers (i.e., low-probability regions) much higher than thermal energy ($\gg\kT$). This leads to the kinetic entrapment of the system in a single metastable state, as on the timescales obtained in standard atomistic simulations, transitions to other metastable states are infrequent events. Such transitions between metastable states can be related to a few slow degrees of freedom that define a low-dimensional energy landscape. Examples of processes exhibiting metastability include catalysis~\cite{piccini2022ab}, phase and glass transitions~\cite{berthier2011theoretical,sosso2016crystal,van2021towards}, photoactivation~\cite{bondanza2020molecular,rydzewski2022enhancing}, and ligand dissociation~\cite{rydzewski2017ligand,rydzewski2018kinetics,wang2018frequency,rydzewski2019finding}.

A possible resolution to the sampling problem is given by enhanced sampling methods~\cite{abrams2014enhanced,valsson2016enhancing,bussi2020using,kamenik2021enhanced,henin2022enhanced}. Over the years, various strategies for enhanced sampling have emerged, e.g., tempering, variational, or biasing approaches; see Ref.~\citenum{henin2022enhanced} for classification and references therein. In this article, we consider a class of such enhanced sampling methods based on the work by Torrie and Valleau~\cite{torrie1977nonphysical}, which devised a framework for enhanced sampling that modifies the Boltzmann probability distribution by introducing a bias potential acting in a low-dimensional space of collective variables (CVs) that correspond to slow degrees of freedom. However, identifying the reduced space of these CVs capturing the underlying chemical processes must be done before enhanced sampling simulations; it is far from trivial and often relies on experience and intuition. Consequently, many data-driven approaches are used to perform dimensionality reduction and construct CVs using samples directly from exploratory trajectories~\cite{ferguson2011nonlinear,rohrdanz2013discovering,hashemian2013modeling,rydzewski2016machine,chiavazzo2017intrinsic,zhang2018unfolding,hase2019machine,rydzewski2020multiscale,glielmo2021unsupervised,shires2021visualizing,morishita2021time}.

An example of such data-driven approaches is manifold learning~\cite{izenman2012introduction}. The core of most manifold learning methods is having a notion of similarity between high-dimensional data samples, usually through a distance metric~\cite{belkin2001laplacian,hinton2002stochastic,maaten2008visualizing}. The distances are integrated into a global parametrization of the data using kernels to represent a Markov chain containing information about transition probabilities that can be used to learn a smooth and low-dimensional manifold that captures the essentials of the data. This way, we can employ dimensionality reduction methods to learn CVs corresponding to slow degrees of freedom. We can distinguish two main approaches that manifold learning methods take to obtain a mapping to a low-dimensional representation of data: ($i$) eigendecomposition~\cite{molgedey1994separation,roweis2000nonlinear,tenenbaum2000global,belkin2003laplacian,coifman2005geometric,nadler2006diffusion,coifman2006diffusion,coifman2008diffusion,tiwary2016spectral} and ($ii$) divergence optimization~\cite{hinton2002stochastic,maaten2008visualizing,mcinnes2018umap}.

When using manifold learning on dynamical data resulting from atomistic simulations, these data must contain statistically sufficient information about the sampled chemical process. If a high-dimensional data set used in manifold learning does not capture the rare transitions between metastable states, the learned low-dimensional CVs will neither. Unbiased atomistic simulations by construction sample only a fraction of the available configuration space and generally capture fast equilibrium processes. Therefore, employing such unbiased simulations as learning data sets for manifold learning methods can lead to undersampled and non-optimal CVs that do not capture the slow degrees of freedom corresponding to the rare chemical processes.

We can circumvent this issue by using learning data set from enhanced sampling simulations where transitions between metastable states are more frequently observed and are no longer rare events. However, in this case, the simulation data set is biased and does not correspond to the real system, as it is sampled from a biased probability distribution. Using these biased simulation data directly in manifold learning algorithms renders low-dimensional manifolds that are also biased (i.e., their geometry, density, and importance) and thus CVs that do not correspond to the chemical process. Therefore, in manifold learning, we need to correctly take into account that we use biased simulation data when learning CVs from enhanced sampling simulations. Despite several attempts in this direction~\cite{ferguson2011integrating,ceriotti2011simplifying,zheng2013rapid,zhang2018unfolding,banisch2020diffusion,trstanova2020local,rydzewski2020multiscale}, this area remains unexplored.

In this work, we consider the problem of using manifold learning methods on data from enhanced sampling simulations. We provide a unified framework for manifold learning to construct CVs using biased simulation data, which we call \emph{reweighted manifold learning}. To this aim, we derive a pairwise reweighting procedure inspired by anisotropic diffusion maps, which accounts for sampling from a biased probability distribution. We term this procedure \emph{diffusion reweighting}. Our framework considers the underlying geometry, density, and importance of the simulation data to construct a low-dimensional manifold for CVs encoding the most informative characteristics of high-dimensional dynamics of the atomistic system.

Our general framework can be used in many manifold learning techniques on data from both standard and enhanced sampling atomistic simulations. We show that our diffusion reweighting procedure can be employed in manifold learning methods that use both eigendecomposition or divergence optimization. We demonstrate the validity and relevance of our framework on both a simple model potential and high-dimensional atomistic systems.

\section{Theory}
In this section, we introduce the theory behind CVs and enhanced sampling (Sec.~\ref{sec:cvs}), reweighting (Sec.~\ref{sec:rew}), and biased data (Sec.~\ref{sec:data}) that we need to derive diffusion reweighting (Sec.~\ref{sec:diffrew}).

\subsection{Collective Variables}
\label{sec:cvs}
In statistical physics, we consider an $n$-dimensional system specified in complete detail by its configuration variables $\bx \in \mathbb{R}^n$. These configuration variables indicate the microscopic coordinates of the system or any other variables (i.e., functions of the microscopic coordinates) relative to the studied process, e.g., an invariant representation. As a result, such a statistical representation is generally of high dimensionality.

In general, the configuration variables $\bx$ are sampled during a simulation according to some, possibly unknown, high-dimensional probability distribution $P(\bx)$ that has a corresponding energy landscape $U(\bx)$ given by the negative logarithm of the probability distribution and an appropriate energy scale. If $\bx$ consists of the microscopic coordinates, this distribution is known and is the stationary Boltzmann distribution:
\begin{equation}
  \label{eq:boletz}
  P(\bx) = \frac{1}{\mathcal{Z}}\e^{-\beta U(\bx)},
\end{equation}
where $U(\bx)$ is the potential energy function of the system, the canonical partition function is $\mathcal{Z}=\int\dx \e^{-\beta U(\bx)}$, and $\beta^{-1}=\kT$ is the thermal energy with $T$ and $k_\mathrm{B}$ denoting the temperature and Boltzmann's constant, respectively. Without loss of generality, we limit the discussion to the canonical ensemble ($NVT$) here.

The high-dimensional description of the system is very demanding to work with directly; hence, many classical approaches in statistical physics were proposed to introduce a coarse-grained representation, e.g., the Mori--Zwanzig formalism~\cite{zwanzig1961memory,mori1965transport} or Koopman's theory~\cite{brunton2021modern}.

To reduce the dimensionality of the high-dimensional space and obtain a more useful representation with a lower number of degrees of freedom, we map the configuration variables to a limited number of functions of the configuration variables, or so-called CVs. A corresponding target mapping $\xi$ is the following:
\begin{equation}
  \label{eq:xtoz}
  \bx \mapsto \xi(\bx) \equiv \set[\big]{ \xi_k(\bx) }_{k=1}^d,
\end{equation}
where $d$ is the number of CVs ($d\ll n$) and $\{ \xi_k \}$ are CVs.

\begin{figure*}[t]
  \includegraphics[width=0.6\textwidth]{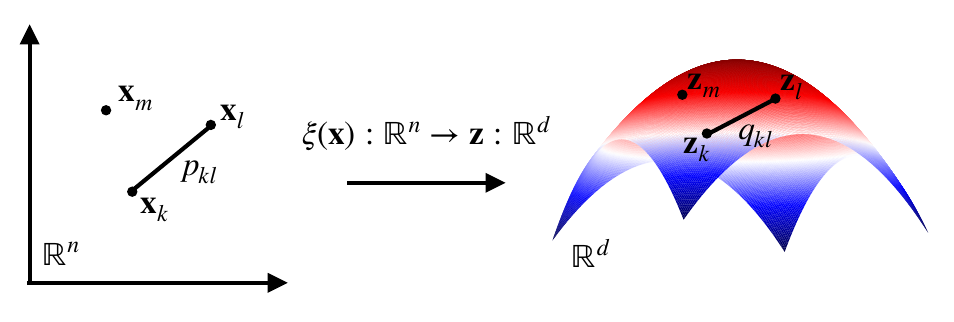}
  \caption{Target mapping from high-dimensional samples of configuration variables $\bx$ to a low-dimensional manifold spanned by CVs $\bz$. In our framework, learning CVs is equivalent to finding the optimal parametrization of the target mapping $\bz=\xi(\bx)$ [\eq{eq:xtoz}]. The target mapping performs the reduction from $\mathbb{R}^n$ to $\mathbb{R}^d$ so the relation $p_{kl}$ between the high-dimensional samples $\bx_k$ and $\bx_l$ is preserved in the relation $q_{kl}$ in a low-dimensional manifold between the CV samples $\bz_k$ and $\bz_l$. For a detailed discussion, see Secs.~\ref{sec:diffrew} and~\ref{sec:rse}.}
  \label{fig:map}
\end{figure*}

The parametrization of the target mapping is performed to retain the system characteristics after embedding into the low-dimensional CV space (Fig.~\ref{fig:map}). In contrast to the configuration variables $\bx$, there are several requirements that the optimal CVs should fulfill, i.e., ($i$) they should be few in number (i.e., the CV space should be low-dimensional), ($ii$) they should correspond to slow modes of the system, and ($iii$) they should separate relevant metastable states. If these requirements are met, we can quantitatively describe rare events.

Let us assume the target mapping and the CVs are known. Then, we can calculate the equilibrium marginal distribution of CVs by integrating over other variables:
\begin{equation}
  \label{eq:cveq}
    P(\bz) = \int\dx \delta\paren*{\bz - \xi(\bx)} P(\bx),
\end{equation}
where the $\delta$-distribution is $\delta(\bz-\xi(\bx))=\prod_k \delta(z_k-\xi_k(\bx))$.

Having the marginal equilibrium probability, we can define the free-energy landscape in the CV space as the negative logarithm multiplied by the thermal energy:
\begin{equation}
  F(\bz)=-\frac{1}{\beta}\log P(\bz).
\end{equation}

In practice, free-energy landscapes for systems severely affected by the sampling problem are characterized by many metastable states separated by high kinetic barriers that impede transitions between metastable states. Consequently, on the timescales we can simulate, the system stays kinetically trapped in a single free-energy minimum and cannot explore the CV space efficiently.

\subsection{Enhanced Sampling}
\label{sec:es}
CV-based enhanced sampling techniques overcome the sampling problem by introducing a bias potential $V(\bz)$ acting in the CV space designed to enhance CV fluctuations. The functional form of the bias depends on the enhanced sampling method used~\cite{torrie1977nonphysical,laio2002escaping,barducci2008well,valsson2014variational,valsson2016enhancing,henin2022enhanced}. The bias potential can be static~\cite{torrie1977nonphysical} or adaptively constructed on the fly during the simulation~\cite{laio2002escaping,barducci2008well,valsson2014variational,valsson2016enhancing,henin2022enhanced}. Regardless of how the bias potential is constructed, it leads to a biased CV distribution that is smoother and easier to sample than the unbiased distribution [\eq{eq:cveq}]:
\begin{align}
  \label{eq:cvbi}
  P_V(\bz) = \ang[\Big]{\delta\paren*{\bz - \xi(\bx)}}_V
           = \frac{1}{\mathcal{Z}_V} \e^{-\beta \paren*{F(\bz)+V(\bz)}},
\end{align}
where $\ang*{\cdot}_V$ denotes the biased ensemble average and the biased partition function is $\mathcal{Z}_V=\int\dz \e^{-\beta \paren*{F(\bz)+V(\bz)}}$.

CV-based enhanced sampling methods construct the bias potential to reduce or entirely flatten free-energy barriers. Let us consider well-tempered metadynamics~\cite{barducci2008well}, which is the method we employ in this work. Well-tempered metadynamics uses a history-dependent bias potential updated iteratively by periodically depositing Gaussians centered at the current location in the CV space. The bias potential is given as:
\begin{equation}
  \label{eq:bias}
  V(\bz) = \sum_l G_{\sigma}(\bz,\bz_l) \exp\paren*{-\frac{1}{\gamma-1}\beta V(\bz_l)},
\end{equation}
where $G_{\sigma}(\bz,\bz_l)$ is a Gaussian kernel with a bandwidth set $\sigma$, $\bz_l$ is the center of $l$-th added Gaussian, and $\gamma$ is a bias factor that determines how much we enhance CV fluctuations. Well-tempered metadynamics convergences to a biased CV distribution given by the so-called well-tempered distribution:
\begin{equation}
\label{eq:wt-probdist}
  P_V(\bz) =
    \frac{\bkt*{P(\bz)}^{1/\gamma}}
    {\int \dz \, \bkt*{P(\bz)}^{1/\gamma}},
\end{equation}
which we can view as sampling an effective free-energy landscape $F/\gamma$ with barriers reduced by a factor of $\gamma$.

\subsection{Reweighting}
\label{sec:rew}
Biasing results in gradual diverging from the equilibrium CV distribution to a smoother and easier to sample biased CV distribution, i.e., from \eq{eq:cveq} to \eq{eq:wt-probdist} in the case of well-tempered metadynamics. Consequently, the importance of each sample is given by a statistical weight needed to account for the effect of the bias potential when obtaining equilibrium properties such as the free-energy landscape. This contrasts with unbiased simulations where samples are equally important as they are sampled according to the equilibrium distribution.

A functional form of the weights depends on a particular method. Generally, for methods employing a bias potential $V(\bz)$, the weight associated with a CV sample $\bz$ can be written as:
\begin{equation}
  \label{eq:weight}
  w(\bz) = \e^{\beta V(\bz)}.
\end{equation}
In the case of a static bias, the weights are given by \eq{eq:weight}. In contrast, well-tempered metadynamics uses an adaptive bias potential [\eq{eq:bias}], and we need to account for a time-dependent constant given by~\cite{tiwary_rewt,valsson2016enhancing}:
\begin{equation}
  \label{eq:coft}
  c = \frac{1}{\beta}\log \frac{\int\dz\exp\paren*{\frac{\gamma}{\gamma-1}\beta V(\bz)}}{\int\dz\exp\paren*{\frac{1}{\gamma-1}\beta V(\bz)}},
\end{equation}
which is independent of $\bz$. We can then redefine the weights as:
\begin{equation}
  \label{eq:wtmweights}
  w(\bz)=\e^{\beta(V(\bz) - c)},
\end{equation}
where $V(\bz) - c$ is called the relative bias potential.

Note that in the above discussion, we assume that the dependence of the bias potential on the simulation time is implicit. We can ignore the time dependence once the simulation reaches convergence, then the relative bias potential $V(\bz) + c$ is quasi-stationary and does not change considerably (the bias potential $V(\bz)$ and the time-dependent constant $c$ can still increase while their sum converges). In practice, when performing reweighting, we ignore a short initial transient part of the simulation where the relative bias potential is still changing considerably.

The standard reweighting works by employing the weights to obtain the stationary equilibrium distribution from the biased CV distribution, i.e., $P(\bz) \propto w(\bz) P_V(\bz)$. The unbiased probability distribution $P(\bz)$ can be computed by histogramming or kernel density estimation, where each sample $\bz$ is weighted by \eq{eq:weight}. This is done routinely in advanced simulation codes, e.g., \textsc{plumed}~\cite{plumed,plumed-nest}.

Manifold learning methods cannot use the standard reweighting to unbias pairwise relations between samples. Instead, a non-trivial approach to reweighting in a form of $r(\bx_k, \bx_l)$ is required, where $r(\bx_k, \bx_l)$ is a pairwise reweighting factor that characterizes the importance of relation between samples $\bx_k$ and $\bx_l$.

\subsection{Biased Data for Manifold Learning}
\label{sec:data}
Given the requirements for the optimal CVs (Sec.~\ref{sec:cvs}), it is non-trivial to provide low-dimensional CVs knowing only the microscopic coordinates. Instead, we often resort to an intermediate description and select a large set of the configuration variables (often called features). For example, this might be internal coordinates such as distances or dihedral angles, and so forth. These configuration variables then define a high-dimensional space which we reduce to the optimal low-dimensional CVs. For a list of helpful configuration variables to characterize different chemical systems, see, for example, the \textsc{plumed} documentation~\cite{plumed-doc}.

Consider data obtained from enhanced sampling simulations in which we record or select samples of the high-dimensional configuration variables $\bx$. These data define the training set from which manifold learning methods construct a low-dimensional manifold. The training data set can be generally expressed as:
\begin{equation}
  \label{eq:data}
  D_V = \set[\Big]{(\bx_k\in \mathbb{R}^n, w(\bx_k))}_{k=1}^K,
\end{equation}
where $K$ is the number of samples and the sample set is augmented by the corresponding statistical weights. Note that the weights depend on $\bx$ through the CV mapping [\eq{eq:xtoz}].

\subsection{Diffusion Reweighting}
\label{sec:diffrew}
Geometrically, the existence of a low-dimensional representation assumes that the high-dimensional dynamical system populates a low-dimensional manifold. This assumption is known as the manifold hypothesis~\cite{ferguson2011integrating}. Under this view, the fast degrees of freedom are adiabatically slaved to the dynamics of the slow degrees of freedom, which correspond to the optimal CVs, due to the presence of fast equilibration within the metastable states. Methods leveraging this assumption belong to a class of manifold learning techniques.

The core of manifold learning methods appropriate for dimensionality reduction in dynamical systems is the construction of a random walk through a Markov chain on the data set, where the transition probabilities $p_{kl}$ depend on a kernel function and distances between samples. Depending on how the transition probabilities $p_{kl}$ are used to find a target mapping to a low-dimensional manifold, we can distinguish two main approaches: ($i$) eigendecomposition~\cite{molgedey1994separation,roweis2000nonlinear,tenenbaum2000global,belkin2003laplacian,coifman2005geometric,nadler2006diffusion,coifman2006diffusion,coifman2008diffusion,tiwary2016spectral} and ($ii$) divergence optimization~\cite{hinton2002stochastic,maaten2008visualizing,mcinnes2018umap}. In manifold learning methods using eigendecomposition, eigenvalues and eigenvectors are used to construct the target mapping. In methods employing divergence optimization, however, the transition probabilities $p_{kl}$ are used to find a Markov transition matrix $q_{kl}$ constructed from low-dimensional samples (Fig.~\ref{fig:map}).

Although many kernels can be considered in manifold learning, a typical choice in spectral embedding methods is a Gaussian kernel dependent on Euclidean distances~\cite{belkin2001laplacian,coifman2005geometric}:
\begin{equation}
  \label{eq:gauss}
  p_{kl} \sim G_\varepsilon(\bx_k, \bx_l) = \exp\paren*{-\frac{1}{\varepsilon}\| \bx_k - \bx_l \|^2},
\end{equation}
where $\varepsilon$ is a positive parameter chosen depending on the given data set as it induces a length scale ${\sim}\sqrt{\varepsilon}$ that should match the distance between neighboring samples. \eq{eq:gauss} models the Markov transition matrix if every row is normalized to unity.

However, this construction includes information only on the manifold geometry given by the pairwise distances. The remaining components required for our reweighting approach are the density and importance of the data.

For the Markov transition matrix, the reweighting procedure must be reformulated to include the weights $w(\bx_k)$ and $w(\bx_l)$ for a pair of samples $\bx_k$ and $\bx_l$, respectively. Our plan is to derive such a pairwise reweighting formula where each pairwise transition probability given by the Markov transition matrix $M(\bx_k,\bx_l)$ depends also on a reweighting factor $r(\bx_k,\bx_l)$. We assume that a reweighted Markov transition matrix can be defined in a simple form:
\begin{equation}
  \label{eq:rgenform}
  M(\bx_k,\bx_l) \propto r(\bx_k,\bx_l) G_\varepsilon(\bx_k,\bx_l),
\end{equation}
where $M$ is row-stochastic. The Markov transition matrix models then the unbiased Markov chain where each entry is the probability of the jump from $\bx_k$ to $\bx_l$.

To account for the manifold density, we need to employ a density-preserving kernel. In contrast to Laplacian eigenmaps that are appropriate for data sampled uniformly~\cite{belkin2001laplacian,belkin2003laplacian}, diffusion map allows working with data sampled from any underlying probability distribution. Specifically, let us consider the pairwise transition probabilities based on an anisotropic diffusion kernel given by~\cite{coifman2005geometric}:
\begin{equation}
  \label{eq:dm}
  \kappa(\bx_k, \bx_l) = \frac{G_\varepsilon(\bx_k, \bx_l)}{[\rho(\bx_k)]^\alpha [\rho(\bx_l)]^\alpha},
\end{equation}
where $\rho(\bx)$ is a kernel density estimator and $\alpha \in [0,1]$ is the anisotropic diffusion parameter, which is crucial to properly include information about the data density and importance~\cite{nadler2006diffusion}. Based on the anisotropic diffusion parameter, diffusion map can be used to parametrize a family of low-dimensional embeddings.

In \eq{eq:dm}, the density estimator $\rho(\bx_k)$ at a sample $\bx_k$ must be reweighted to account on the data importance:
\begin{equation}
  \label{eq:rhop}
  \rho(\bx_k)=\sum_l w(\bx_l) G_\varepsilon(\bx_k,\bx_l),
\end{equation}
which is a weighed kernel density estimate up to an unimportant multiplicative constant. After the reweighting, the density estimator characterizes the unbiased density, in contrast to the biased density estimate that is given as:
\begin{equation}
  \label{eq:rhobiased}
  \rho_V(\bx_k)=\sum_l G_\varepsilon(\bx_k,\bx_l),
\end{equation}
where the subscript $V$ denotes that the density estimate is calculated under the bias potential $V$.

In theory, if the underlying probability distribution of high-dimensional samples is known analytically, it is possible to express $\rho$ directly from this distribution~\cite{coifman2008diffusion}; e.g., from a Boltzmann distribution [\eq{eq:boletz}] if the samples are represented by the microscopic coordinates. However, this is valid only in the case of sufficient sampling and thus rarely reachable in practice. Moreover, the high-dimensional distribution $P(\bx)$ of the configuration variables is unknown in general (Sec.~\ref{sec:cvs}). For this reason, we write $\rho$ as a kernel density estimate [\eq{eq:rhop}].

We can understand the meaning behind the anisotropic diffusion kernel by considering \eq{eq:dm}. The dynamics described by \eq{eq:dm} is local as samples closer to each other have a higher probability of being close in the respective low-dimensional manifold and vice versa in the case that they are farther apart. This information about the underlying geometry is given by $G_\varepsilon(\bx_k, \bx_l)$ which requires that the transition probabilities are penalized between geometrically distant samples $\bx_k$ and $\bx_l$. The density and importance of samples are encoded in the unbiased density estimates [\eq{eq:rhop}].

Depending on $\alpha$ value in \eq{eq:dm}, three interesting cases of diffusion maps can be considered asymptotically~\cite{nadler2006diffusion}. Namely, ($i$) for $\alpha=\frac{1}{2}$, \eq{eq:dm} corresponds to the Markov chain that is an approximation of the diffusion given by the Fokker-Planck generator with the underlying data density proportional to the equilibrium density, allowing us to approximate the long-time behavior of the microscopic coordinates. Other values of $\alpha$ are also possible, e.g., ($ii$) for $\alpha=0$, we get the classical normalized graph Laplacian, and ($iii$) for $\alpha=1$, we ignore the underlying density and the diffusion operator approximates the Laplace-Beltrami operator. We note that this asymptotic behavior holds in the limit of infinite data $K\rightarrow\infty$ and $\varepsilon\rightarrow 0$ when considering the microscopic coordinates. As we are interested in finding low-dimensional CVs, the case for $\alpha=\frac{1}{2}$ is appropriate to model asymptotically the slowest degrees of freedom, accounting for both the underlying geometry and density of the manifold.

As we have all the required ingredients for the reweighting of Markov transition matrices, we focus on deriving the reweighting factor. Here, we discuss only an outline, while a detailed derivation is provided in Appendix~\ref{app:diffrew}.

Based on \eq{eq:dm}, the Markov transition matrix can be estimated by weighting each Gaussian term and normalizing it so that it is row-stochastic:
\begin{equation}
  \label{eq:markov}
  M(\bx_k, \bx_l) = \frac{w(\bx_l) \kappa(\bx_k,\bx_l)}{\sum_m w(\bx_m) \kappa(\bx_k,\bx_m)}.
\end{equation}
Next, by inserting \eq{eq:dm} to \eq{eq:markov}, we can see that the Markov transition matrix $M$ can be written also using the Gaussian kernels:
\begin{equation}
  \label{eq:markovgen}
    M(\bx_k, \bx_l) \propto \frac{w(\bx_k)}{{[\rho(\bx_k)]^\alpha}}
    \frac{w(\bx_l)}{{[\rho(\bx_l)]^\alpha}}
    G_\varepsilon(\bx_k, \bx_l),
\end{equation}
where we can recognize the reweighting factor by comparing the result to \eq{eq:rgenform}. Therefore, we get the following expression:
\begin{align}
  \label{eq:rfactor}
  r(\bx_k,\bx_l) = \frac{w(\bx_k)}{{[\rho(\bx_k)]^\alpha}} \frac{w(\bx_l)}{{[\rho(\bx_l)]^\alpha}}.
\end{align}

We can also approximate the reweighting factor by rewriting \eq{eq:rfactor} with the biased density estimate [\eq{eq:rhobiased}]:
\begin{align}
  \label{eq:rfactorslow}
  r(\bx_k,\bx_l) &= \frac{w(\bx_k)}{\sqrt{\rho(\bx_k)}} \frac{w(\bx_l)}{\sqrt{\rho(\bx_l)}} \nonumber\\
  &\approx
  \frac{w(\bx_k)}{\sqrt{w(\bx_k) \rho_V(\bx_k)}}
  \frac{w(\bx_l)}{\sqrt{w(\bx_l) \rho_V(\bx_l)}} \nonumber\\
  &=
    \sqrt{\frac{w(\bx_k)}{\rho_V(\bx_k)}}
    \sqrt{\frac{w(\bx_l)}{\rho_V(\bx_l)}},
\end{align}
where we set $\alpha=\frac{1}{2}$. \eq{eq:rfactorslow} is a final form of the reweighting factor that we use here. A detailed derivation of \eq{eq:rfactorslow} is provided in Appendix~\ref{app:diffrew}. Although the derivation of \eq{eq:rfactorslow} is presented using the Gaussian kernel, our framework can be used in other manifold learning methods, as demonstrated in Sec.~\ref{sec:rml}.

\eq{eq:markovgen} denotes an unbiased Markov chain with the transition probability from $\bx_k$ to $\bx_l$ in one time step $t$ given by:
\begin{equation}
  \label{eq:prob}
  \mathrm{Pr}\set[\big]{\bx(t+1)=\bx_l\,|\,\bx(t)=\bx_k} = M(\bx_k, \bx_l).
\end{equation}

We term our reweighting procedure diffusion reweighting. We postulate that the derived Markov transition matrix [\eq{eq:markovgen}] has the following three properties that make the construction of \eq{eq:prob} from enhanced sampling simulations feasible. Namely, the Markov transition matrix encodes the information about:
\begin{enumerate}[leftmargin=0.6cm]
  \item Geometry $G_\varepsilon(\bx_k,\bx_l)$: The probability of transitions between samples lying far from each other is low and high for those in close proximity.
  \item Density $[\rho(\bx_l)]^\alpha$: The anisotropic diffusion constant $\alpha \in [0,1]$ is used as a density-scaling term as in diffusion maps. See \eq{eq:dm} and the corresponding description.
  \item Importance $w(\bx_l)$: The statistical weights from enhanced sampling decide accordingly to the bias if a sample is important, i.e., metastable states where the weights are higher are more important then high free-energy regions.
\end{enumerate}

\subsection{Implementation}
Our framework is implemented in a development version of \textsc{plumed} 2.7~\cite{plumed,plumed-nest} as the LowLearner module and will be made publicly available in the coming future. Its initial implementation incorporating several algorithms used in this work can be accessed at Zenodo (\textsc{doi}: \href{https://zenodo.org/record/4756093}{\tt 10.5281/zenodo.4756093}) and from \textsc{plumed-nest}~\cite{plumed-nest} repository under {\tt plumID:21.023} at \url{https://plumed-nest.org/eggs/21/023/}.

\section{Reweighted Manifold Learning}
\label{sec:rml}
We incorporate diffusion reweighting into several manifold learning methods and apply them to find a low-dimensional representation in a model system and high-dimensional atomistic simulation problems represented by biased simulation data. Specifically, we consider diffusion reweighting in diffusion maps~\cite{nadler2006diffusion,coifman2006diffusion,coifman2008diffusion} and recently introduced stochastic embedding methods for learning CVs and adaptive biasing~\cite{zhang2018unfolding,rydzewski2020multiscale}.

To demonstrate the validity of our framework, we apply diffusion map to standard testing systems such as a particle moving on an analytical potential and alanine dipeptide. For the stochastic embedding methods, we choose a mini-protein chignolin. For the two atomistic systems, alanine dipeptide and chignolin, we describe the systems using two different types of high-dimensional representations (distances and dihedral angles, respectively) to show that the framework can work regardless of the chosen configuration variables.

\subsection{Diffusion Maps}
We start by considering the case of diffusion maps on which we base the derivation of the reweighting factor $r(\bx_k, \bx_l)$ (Sec.~\ref{sec:diffrew}). By rewriting the diffusion kernel using the biased density estimates [\eq{eq:rfactorslow}], we can use it to construct a low-dimensional embedding from a biased data set. We directly use \eq{eq:markovgen} to estimate the transition probabilities while using \eq{eq:rfactorslow} to account for the sampling from any biased distribution.

\subsubsection{Target Mapping $\xi(\bx)$: Eigendecomposition}
With the exemption of the reweighting factor, further steps in our approach to diffusion maps proceed as in its standard formulation~\cite{nadler2006diffusion}. Let us briefly recap these steps.

In diffusion maps, the spectral decomposition of the Markov transition matrix $M$ is performed to define a low-dimensional embedding, $M\psi=\lambda\psi$, where $\set{\lambda_l}$ and $\set{\psi_l}$ are eigenvalues and eigenvectors, respectively. The eigenvalues are related to the effective timescales as $\tau_l = -\frac{1}{\log \lambda_l}$ and can be used to determine the slowest processes in the dynamics. Then, the eigenvectors corresponding to the largest eigenvalues define a reduced space. Given this interpretation, the target mapping [\eq{eq:xtoz}] is defined by the diffusion coordinates:
\begin{equation}
  \label{eq:diffcoords}
  \xi(\bx) = \set[\big]{\lambda_k\psi_k(\bx)}_{k=0}^{d-1},
\end{equation}
where $\xi(\bx)$ is computed using the first $d$ eigenvalues and eigenvectors with the the equilibrium density represented by the zeroth coordinate $\lambda_0\psi_0$. In \eq{eq:diffcoords}, the spectrum of the eigenvalues $\set{\lambda_l}$ is sorted by non-increasing value, $\lambda_0 = 1 > \lambda_1 \ge \dots \ge \lambda_{d-1}$.

The truncation up to $d-1$ of \eq{eq:diffcoords} for metastable systems corresponds to a negligible error on the order of $O(\lambda_{d}/\lambda_{d-1})$~\cite{nadler2006diffusion}. In other words, this assumption relates to a large spectral gap that separates slow degrees of freedom ($>\lambda_{d-1}$) and fast degrees of freedom ($<\lambda_{d-1}$). For a detailed description behind the construction of the diffusion coordinates from unbiased data, we refer to works by Coifman~\cite{coifman2005geometric,coifman2006diffusion,coifman2008diffusion}. By inspecting the spectral gap obtained via the eigendecomposition of the reweighted Markov transition matrix, it is possible to verify that the selected high-dimensional representation sampled from a biased distribution contains enough information to render a physically-meaningful low-dimensional manifold.

\subsubsection{Algorithm}
The described algorithm for our reweighted diffusion maps is given in Algorithm~\ref{alg:dm}.
\begin{algorithm}
  \SetKwInOut{Input}{Input}
  \SetKwInOut{Output}{Output}
  \Input{Biased data set $\set[\big]{(\bx_k, w(\bx_k))}_{k=1}^K$.}
  \Output{Eigenvalues $\set{\lambda_k}$ and eigenvectors $\set{\psi_k}$ of the transition matrix $M$ used to construct the target mapping $\xi$.}
  \begin{enumerate}[leftmargin=0cm]
    \item Calculate the squared pairwise distances $\| \bx_l - \bx_l \|^2$.
    \item Estimate the Markov transition matrix $M(\bx_k,\bx_l)$:
    \begin{enumerate}[leftmargin=0.6cm]
      \item Apply the Gaussian kernel $G_\varepsilon(\bx_k,\bx_l)$ [\eq{eq:gauss}].
      \item Construct the anisotropic diffusion kernel using the reweighting factor $r(\bx_k,\bx_l)$ [\eq{eq:rfactorslow}].
      \item Normalize to obtain the row-stochastic matrix $M$.
    \end{enumerate}
    \item Perform eigendecomposition $M\psi=\lambda\psi$ and estimate the diffusion coordinates $\xi(\bx)$ [\eq{eq:diffcoords}].
  \end{enumerate}
  \caption{Reweighted Diffusion Maps}
  \label{alg:dm}
\end{algorithm}

\subsubsection{Example: Model Potential}
\begin{figure*}[t]
  \includegraphics{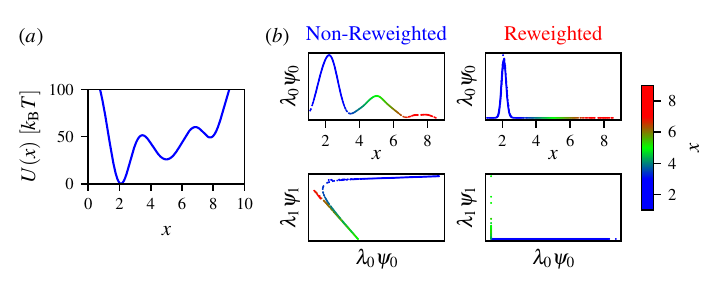}
  \caption{Diffusion maps generated for the reweighted and non-reweighted (without applying diffusion reweighting) biased simulation of a particle in a simple ($a$) one-dimensional potential $U(x)$ where the energy barriers separating the deepest minimum are on the order of 50 $\kT$, and the corresponding transitions from this state are rare events. ($b$) A comparison between the non-reweighted (blue) and reweighted (red) diffusion maps: the equilibrium densities along the coordinate $x$ and diffusion coordinates $\lambda_0\psi_0$ vs. $\lambda_1\psi_1$, with coloring according to the $x$ value. The enhanced sampling simulation is performed using well-tempered metadynamics~\cite{barducci2008well} with a bias factor of 10 by employing the \texttt{pesmd} code in the \textsc{plumed}~\cite{plumed,plumed-nest} plugin.}
  \label{fig:pot}
\end{figure*}

As a simple and illustrative example of applying diffusion reweighting within the diffusion map framework, we consider a case where dimensionality reduction is not performed. Namely, we run an enhanced sampling simulation of a single particle moving along the $x$ variable on a one-dimensional potential $U(x)$ with three Gaussian-like metastable states with different energy depths and energy barriers between the minima [Fig.~\ref{fig:pot}($a$)]. In this system, the highest energy barrier is ${\sim}50~\kT$, which makes the transitions from the deepest minimum rare. The dynamics is modeled by a Langevin integrator~\cite{bussi2007accurate} using temperature $T=1$, a friction coefficient of 10, and a time step of 0.005. We employ the \texttt{pesmd} code in the \textsc{plumed}~\cite{plumed,plumed-nest} plugin. We bias the $x$ variable using well-tempered metadynamics~\cite{barducci2008well} with a bias factor of $\gamma=10$. Further details about the simulation are given in Supporting Information (SI) in Sec. S1 A.

We present our results in Fig.~\ref{fig:pot}($b$). We can see that the non-reweighted (without applying diffusion reweighting) diffusion map learns the biased distribution (given by $\lambda_0\psi_0$) along the coordinate $x$ where the three energy minima correspond to the maxima of the biased distribution. Additionally, the first two diffusion coordinates are not orthogonal, and there is a lack of separation between the metastable states.

In contrast, the reweighted diffusion map can represent the equilibrium density ($\lambda_0\psi_0$) where only the first energy minimum is populated due to the high-free energies separating the states. The $\lambda_0\psi_0$ and $\lambda_1\psi_1$ diffusion coordinates properly separate the samples. We can see that $\lambda_1\psi_1$ is almost marginal due to the lack of additional dimensions for the potential energy.

The example presented in Fig.~\ref{fig:pot} is, of course, a trivial case in which no dimensionality reduction is performed; however, it indicates that diffusion reweighting can be used to reweight the transition probabilities successfully and that the standard diffusion map trained on the biased data captures an incorrect representation.

\subsubsection{Example: Alanine Dipeptide}
\label{sec:diala}
\begin{figure*}[t]
  \includegraphics{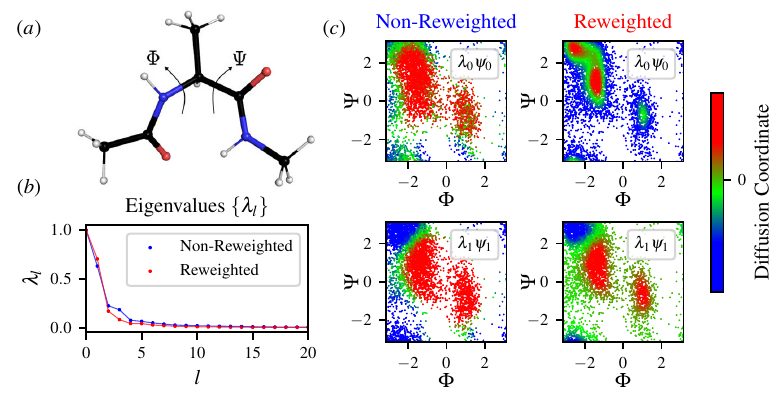}
  \caption{Reweighted diffusion maps on a peptide model system (Ace-Ala-Nme) in vacuum at 300 K simulated using well-tempered metadynamics~\cite{barducci2008well} enhancing the $\Phi$ and $\Psi$ dihedral angles and a bias factor $\gamma=5$. The diffusion map is calculated using a high-dimensional space of 45 pairwise distances between heavy atoms. ($a$) A representative structure of alanine dipeptide with the dihedral angles $\Phi$ and $\Psi$. ($b$) A spectrum of eigenvalues $\{\lambda_l\}$ obtained from the eigendecomposition for the non-reweighted (blue) and reweighted (red) Markov transition matrices. ($c$) The samples are shown in the dihedral angle space for the non-reweighted (blue label) and reweighted (red label) diffusion map with colors representing the first and second diffusion-map coordinates $\lambda_0\psi_0(\bx)$ and $\lambda_1\psi_1(\bx)$, respectively. The color bar represents the constructed diffusion coordinates.}
  \label{fig:ala}
\end{figure*}

As a next example, we consider alanine dipeptide (Ace-Ala-Nme) in gas phase described using the Amber99-SB force field~\cite{hornak2006comparison}. The data set is generated by a 100-ns molecular dynamics simulation~\cite{vrescale,hess2008p} using the \textsc{gromacs} 2019.2 code~\cite{gromacs} patched with a development version of the \textsc{plumed}~\cite{plumed,plumed-nest} plugin. The simulation is performed by well-tempered metadynamics~\cite{barducci2008well} at 300 K using the backbone dihedral angles $\Phi$ and $\Psi$ for biasing and a bias factor of 5. We select the $\Phi$ and $\Psi$ dihedral angles as biasing them is sufficient to sample accelerated transitions between several metastable states of alanine dipeptide. Using this setup, the convergence of the bias potential is obtained quickly. Further details about the simulation are given in SI (Sec. S1 B).

Using diffusion maps, we reduce the high-dimensional space consisting of all pairwise distances between the heavy atoms ($n=45$) to two dimensions. The diffusion maps are constructed using $\varepsilon=0.078$ estimated as the median of the pairwise distances.

We present diffusion reweighting results for alanine dipeptide in Fig.~\ref{fig:ala}. The eigenvalues of the Markov transition matrix have a spectral gap (i.e., timescale separation) with only a few eigenvalues close to one and all other eigenvalues much smaller than one. Thus, only the first few eigenvectors are needed to approximate the diffusion coordinates [\eq{eq:diffcoords}] and thus the target mapping to the CV space. The eigenvalues $\{\lambda_l\}$ indicate that the spectral gap is slightly wider for the reweighted transition probability matrix, as can be seen in Fig.~\ref{fig:ala}($b$). Consequently, the effective timescales $\tau_l = -\frac{1}{\log \lambda_l}$ calculated from the eigenvalues indicate that the reweighted diffusion map corresponds to slower processes; see SI (Figs. S3 and S4).

We can see that the non-reweighted approach cannot correctly account for the transition probabilities calculated based on the biased simulation, as we expected. The transitions between the metastable states are so frequent that the first diffusion coordinate (the equilibrium density) suggests only one metastable state [Fig.~\ref{fig:ala}($c$)]. In SI (Fig. S2), we show that the separation of samples in the reweighted diffusion map is much better than for the non-reweighted diffusion map. It resembles a ``typical'' diffusion map from unbiased data sets.

In the reweighted case, the low-dimensional coordinates can distinguish between the relevant metastable states. Additionally, using \eq{eq:rfactorslow} the first diffusion-map coordinate, $\lambda_0\psi_0(\bx)$, correctly encodes the information about the Boltzmann equilibrium distribution of alanine dipeptide in the dihedral angle space, which is not possible using the standard (i.e., non-reweighted) diffusion map in the case of biased simulation data [Fig.~\ref{fig:ala}($c$)]. By comparing the reweighted diffusion map to a diffusion map constructed from an unbiased parallel tempering replica at 300 K, we can see that the embeddings and eigenvalues are virtually identical; see SI (Fig. S5).

These results further corroborate our findings and show that when performing a dimensionality reduction from data resulting from enhanced sampling, the reweighting factor [\eq{eq:rfactorslow}] is needed to revert the effect of biasing in the transition probability matrix.

\subsection{Stochastic Embeddings}
\label{sec:rse}
Next, we move to employ diffusion reweighting in more recent approaches. We consider manifold learning methods devised primarily to learn CVs from biased simulation trajectories: multiscale reweighted stochastic embedding (\textsc{mrse})~\cite{rydzewski2020multiscale} and stochastic kinetic embedding (\textsc{stke})~\cite{zhang2018unfolding}. These methods use approximations of the reweighting factor [\eq{eq:rfactorslow}]. Our aim is not to compare results obtained using these methods but to present and discuss how diffusion reweighting can be approximated and employed in manifold learning methods other than diffusion maps.

First, let us focus on a general procedure these stochastic embedding methods use to parametrize manifolds. Mainly, we discuss how these methods use the Markov transition matrices to parametrize the target mapping to low-dimensional manifolds. The construction of the Markov transition matrix with reweighting from biased data in each technique is discussed in the remainder of this section.

\subsubsection{Target Mapping $\xi_{\boldsymbol{\theta}}(\bx)$: Divergence Optimization}
As mentioned above, the stochastic embedding methods belong to the second category of manifold learning methods we consider here, i.e., based on divergence optimization. Thus, unlike diffusion maps, the eigendecomposition is not performed in these methods. Instead, the target mapping $\xi$ is parametrized based on neural networks that perform nonlinear dimensionality reduction. The target mapping is given as:
\begin{equation}
  \label{eq:map}
  \bz:\bx \mapsto \xi_{\boldsymbol{\theta}}(\bx),
\end{equation}
where $\boldsymbol{\theta}=\set{\theta_k}$ are parameters of the target mapping adjusted such that the low-dimensional manifold of CVs is optimal with respect to a selected statistical measure. Using \eq{eq:map}, the distance between samples in a manifold can be given as:
\begin{equation}
  \|\bz_k - \bz_l\| = \| \xi_{\boldsymbol{\theta}}(\bx_k) - \xi_{\boldsymbol{\theta}}(\bx_l) \|.
\end{equation}

Note that in some simple cases, the mapping in \eq{eq:map} can also be represented using a linear combination. However, deep learning has been successful in a broad range of learning problems, and using more intricate approximations for the mapping between high-dimensional and low-dimensional spaces is quite common for complex data sets~\cite{hinton2006reducing,maaten2009learning}.

The target mapping is parametrized by comparing the Markov transition matrix $M(\bx_k,\bx_l)=(p_{kl})$ (Sec.~\ref{sec:rmt}) constructed from the high-dimensional samples to a Markov transition matrix $Q(\bz_k,\bz_l)=(q_{kl})$ built from low-dimensional samples mapped using the target mapping [\eq{eq:map}].

In \textsc{stke}, we use a Gaussian kernel for $Q$:
\begin{align}
  \label{eq:qgauss}
  q_{kl} \sim \exp\paren*{-\frac{1}{\varepsilon}\|\xi_{\boldsymbol{\theta}}(\bx_k) - \xi_{\boldsymbol{\theta}}(\bx_l) \|^2}.
\end{align}

In \textsc{mrse}, we employ a one-dimensional $t$-distribution, as implemented in $t$-\textsc{sne}~\cite{maaten2008visualizing,maaten2009learning}. Taking the target mapping as defined in \eq{eq:map}, the transition probabilities in the low-dimensional space $q_{kl}$ in \textsc{mrse} are:
\begin{align}
  \label{eq:tdist}
  q_{kl} \sim \paren[\big]{1+\|\xi_{\boldsymbol{\theta}}(\bx_k) - \xi_{\boldsymbol{\theta}}(\bx_l) \|^2}^{-1}.
\end{align}

The choice of the $t$-distribution for $Q$ in \textsc{mrse} is motivated by the apparent crowding problem~\cite{maaten2008visualizing}, i.e., as the volume of a small-dimensional neighborhood grows slower than the volume of a high-dimensional one, the neighborhood is stretched so that moderately distant sample pairs are placed too far apart. As outlined in Ref.~\citenum{maaten2008visualizing}, the use of a heavy-tailed distribution for the low-dimensional representation allows moderate distances in the high-dimensional space to be represented by much larger distances in the manifold, encouraging gaps to form in the low-dimensional map between the clusters present in the data, alleviating the crowding problem to some degree.

Finally, the Markov transition matrices computed from the high-dimensional and low-dimensional samples need to be compared. The most common choice for such a metric is employing a statistic distance, particularly the Kullback--Leibler divergence:
\begin{equation}
  \label{eq:dkl}
  D_{\mathrm{KL}}(M, Q; \boldsymbol{\theta})=\sum_k \sum_l p_{kl}\log\paren*{\frac{p_{kl}}{q_{kl}}},
\end{equation}
where in contrast to the standard formulation of the Kullback--Leibler divergence that compares two probability distributions, \eq{eq:dkl} is computed for every pair of rows from $M$ and $Q$, and then summed. Equivalently, we can minimize the cross-entropy:
\begin{equation}
  D_{\mathrm{CE}}(M, Q; \boldsymbol{\theta})=-\sum_k \sum_l p_{kl}\log\paren{q_{kl}},
\end{equation}
as the probabilities $p_{kl}$ stay constant during the optimization. There are many choices possible for the comparison between $M$ and $Q$, e.g., the Jensen--Shannon divergence~\cite{zhang2018unfolding}.

The Kullback--Leibler divergence optimization is performed to train the target mapping represented by a neural network. As the target mapping is parametric, the gradients of $D_{\mathrm{KL}}$ with respect to the parameters $\boldsymbol{\theta}=\set{\theta_k}$ of the neural network can be estimated effortlessly using backpropagation. For further details about training neural networks, we refer to Appendix~\ref{app:comparison}.

\subsubsection{Reweighted Markov Transitions}
\label{sec:rmt}
After explaining how the parametric mapping is constructed in the reweighted stochastic embeddings, we proceed to formulate the Markov transition matrices and the reweighting factors for these methods.

First, let us consider the reweighting performed in \textsc{mrse}~\cite{rydzewski2020multiscale}. This method employs the following reweighting factor:
\begin{equation}
  \label{eq:rmrse}
  r(\bx_k,\bx_l) = \sqrt{w(\bx_k)w(\bx_l)},
\end{equation}
where we neglect the biased density estimates $\rho_V$ [cf. \eq{eq:rmrse} and \eq{eq:rfactorslow}]. The reweighting factor [\eq{eq:rmrse}] written as a geometric mean between two statistical weights can be justified by the fact that the bias potential is additive, as shown in \eq{eq:weight}, and a geometric mean is appropriate to preserve this relation. We note that similar reweighting procedures have been used in Refs.~\citenum{zheng2013molecular,banisch2020diffusion,trstanova2020local}.

The Markov transition matrix in \textsc{mrse} is expressed as a Gaussian mixture, where each Gaussian is evaluated for different $\varepsilon$ values and reweighted using \eq{eq:rmrse}:
\begin{equation}
  \label{eq:mmrse}
  M(\bx_k,\bx_l) \propto \sum_{\set{\varepsilon}} \sqrt{w(\bx_l)} G_\varepsilon(\bx_k,\bx_l),
\end{equation}
where we omit the normalization constant for brevity. The sum in \eq{eq:mmrse} is over bandwidths that are automatically estimated and selected to fit that data. Note that many methods can be used for this purpose; however, to facilitate analysis, we use a method from Ref.~\citenum{rydzewski2020multiscale}. As this procedure is mostly technical, for details about estimating bandwidths and constructing the Gaussian mixture, we refer to Appendix~\ref{app:gaussmix}.

Second, let us consider \textsc{stke}. Suppose high-dimensional samples are resampled so that each sample keeps a certain distance away from the others. In that case, the distribution of samples can be viewed as approximately uniform. Then, $w(\bx)$ can be replaced by the unbiased probability density estimator $\rho(\bx)$ in \eq{eq:rmrse}. Thus, the reweighting factor is given by:
\begin{equation}
  \label{eq:stke}
  r(\bx_k,\bx_l) = \sqrt{\rho(\bx_k)\rho(\bx_l)},
\end{equation}
which is the formula used in \textsc{stke}~\cite{zhang2018unfolding,chen2021collective}. The corresponding Markov transition matrix is:
\begin{equation}
  \label{eq:mstke}
  M(\bx_k, \bx_l) \propto \sqrt{\rho(\bx_l)} G_\varepsilon(\bx_k,\bx_l),
\end{equation}
where, as in \eq{eq:mmrse}, the $k$-th reweighting term is canceled out during the normalization.

An interesting property of the transition probabilities used by this method is that by taking an approximation to the normalization constant (Appendix~\ref{app:sqra}), we arrive at a transition probability matrix of similar form as in the square-root approximation of the infinitesimal generator of the Fokker-Planck operator~\cite{lie2013square,heida2018convergences,donati2018estimation,kieninger2020dynamical}:
\begin{equation}
  \label{eq:entropy}
  M(\bx_k, \bx_l) = \sqrt{\frac{\rho(\bx_l)}{\rho(\bx_k)}} G_\varepsilon(\bx_k,\bx_l),
\end{equation}
for a single $\varepsilon$. The square-root approximation has been initially derived by discretizing a one-dimensional Smoluchowski equation~\cite{bicout1998electron}. It can also be shown that \eq{eq:entropy} can be obtained using the maximum path entropy approach~\cite{dixit2019introducing,ghosh2020maximum}.

As many algorithmic choices are available for each procedure incorporated in the reweighted stochastic embedding framework, it is difficult to directly compare \textsc{mrse} and \textsc{stke}. However, we aim to discuss how approximations of the reweighting factor are employed in these methods and how they can be used to learn CVs from biased data. Thus, in the above discussion, we focus on the reweighting procedures for the Markov transition matrices used by these methods. To compare the parameters used by these methods, see Appendix~\ref{app:comparison}.

\subsubsection{Algorithm}
For a general algorithm used by the stochastic embedding techniques to find a low-dimensional manifold of data, see Algorithm~\ref{alg:se}.
\begin{algorithm}
  \SetKwInOut{Input}{Input}
  \SetKwInOut{Output}{Output}
  \Input{Biased data set $\set[\big]{(\bx_k, w(\bx_k))}_{k=1}^K$.}
  \Output{Target mapping $\bx \mapsto \xi(\bx)$ and the CVs $\bz$.}
  \begin{enumerate}[leftmargin=0cm]
    \item Calculate the squared pairwise distances $\| \bx_l - \bx_l \|^2$.
    \item Estimate the Markov transition matrix $M(\bx_k,\bx_l)$ according to the method used and reweight $M$ using the approximation of the reweighting factor (Sec.~\ref{sec:rmt}).
    \item Use the target function $\xi_{\boldsymbol{\theta}}(\bx)$ to estimate parameters $\boldsymbol{\theta} = \set{\theta_k}$:
    \begin{enumerate}[leftmargin=0.6cm]
      \item Compute the transition matrix $Q(\bz_k,\bz_l)=\paren{q_{kl}}$ from the low-dimensional samples using the $t$-distribution as in \textsc{mrse} [\eq{eq:tdist}] or the Gaussian kernel as in \textsc{stke} [\eq{eq:qgauss}].
      \item Use the Kullback-Leibler divergence $D_{\mathrm{KL}}(\boldsymbol{\theta})$ to estimate statistical distance between $Q$ and $M$ given the current parameters $\boldsymbol{\theta}$ [\eq{eq:dkl}].
      \item Repeat until convergence reached.
    \end{enumerate}
    \item Map the high-dimensional samples to CVs using $\xi(\bx)$, where optimal parameters are given by $\argmin D_{\mathrm{KL}}(\boldsymbol{\theta})$.
  \end{enumerate}
  \caption{Reweighted Stochastic Embedding}
  \label{alg:se}
\end{algorithm}

\subsubsection{Example: Chignolin}
\label{sec:cln025}
As an example for the two stochastic embedding methods \textsc{mrse} and \textsc{stke}, we consider folding and unfolding of a ten amino-acid miniprotein chignolin (CLN025)~\cite{honda2008crystal} in the solvent. We employ the CHARMM27 force field~\cite{charmm} and the TIP3P water model~\cite{tip3p}, and we perform the molecular dynamics simulation~\cite{vrescale,hess2008p} using the \textsc{gromacs} 2019.2 code~\cite{gromacs} patched with a development version of the \textsc{plumed}~\cite{plumed,plumed-nest} plugin. Our simulations are performed at 340 K for easy comparison with other simulation data, also simulated at 340 K~\cite{lindorff2011fast,palazzesi2017conformational}. We perform a 1-$\mu$s well-tempered metadynamics simulation with a large bias factor of 20. We select a high bias factor to illustrate that our framework is able to learn metastable states in a low-dimensional manifold even when free-energy barriers are virtually flattened, and the system dynamics is close to diffusive at convergence.

\begin{figure*}[t]
  \includegraphics{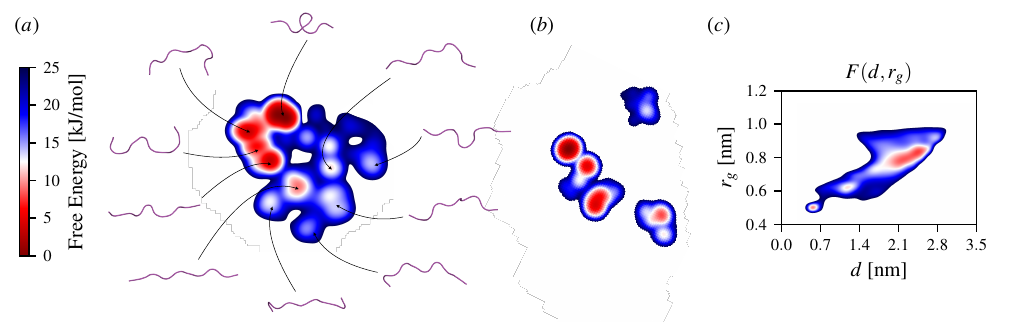}
  \caption{Reweighted stochastic embeddings calculated for chignolin in the TIP3P solvent at 340 K simulated using the CHARMM27 force field. Low-dimensional manifolds are colored according to their free energy. ($a$) Representative conformations from the metastable states estimated by the reweighted embedding methods are shown around the \textsc{mrse} embedding. ($b$) The embedding obtained using \textsc{stke}. Well-tempered metadynamics is used to generate the training set consisting of sines and cosines of all $\Phi$ and $\Psi$ dihedral angles, amounting to 32 variables in total. The training set is generated by performing a 1-$\mu$s simulation with a bias factor $\gamma=20$, enhancing the fluctuations of the distance $d$ between the C$\alpha$ atoms of residues Y1 and Y10 and the radius of gyration $r_g$. ($c$) The free-energy surface calculated along for $d$ and $r_g$. The axes and units for the embeddings are arbitrary and thus not shown. See SI (Sec. S1 C) for computational details.}
  \label{fig:cln}
\end{figure*}

As biased CVs to enhance transitions between the folded and unfolded conformations of CLN025 in the metadynamics simulation, we choose the distance between C$\alpha$ atoms of residues Y1 and Y10 ($d$) and the radius of gyration ($r_g$) [Fig.~\ref{fig:cln}($c$)]. We consider CLN025 conformations folded when the distance is below ${\sim}0.8$ nm and unfolded otherwise for $>0.8$ nm. From the resulting trajectory, we calculate the sines and cosines of all the backbone $\Phi$ and $\Psi$ dihedral angles and use them as the high-dimensional representation of CLN025, which amounts to 32 variables in total. We collect high-dimensional samples every 1 ps for the biased training data set. Then, the low-dimensional manifolds are trained on representative samples selected as described in Refs.~\citenum{zhang2018unfolding,rydzewski2020multiscale}. As we focus mainly on the Markov transition matrices and diffusion reweighting here, we provide a detailed discussion about the subsampling procedures in Appendix~\ref{app:wtrandom}.

In Fig.~\ref{fig:cln}, we present the resulting manifolds spanned by the trained CVs computed using the reweighted stochastic embedding methods (Sec.~\ref{sec:rse}). The embedding presented in Fig.~\ref{fig:cln}($a$) is calculated using \textsc{mrse}~\cite{rydzewski2020multiscale}, while the embedding presented in Fig.~\ref{fig:cln}($b$) is calculated using \textsc{stke}~\cite{zhang2018unfolding}, using their corresponding reweighting formulas given by \eq{eq:rmrse} and \eq{eq:stke}, respectively. For each manifold, the corresponding free-energy landscapes are calculated using kernel density estimation using the weights to reweight each sample [\eq{eq:wtmweights}].

We can observe that the free-energy landscape in the low-dimensional manifold calculated by \textsc{mrse} is highly heterogeneous, with multiple partially unfolded intermediate states and many possible reaction pathways, as shown in Fig.~\ref{fig:cln}($a$). Such a complex free-energy landscape shows that the dynamics of CLN025 is more intricate and complex than it is visible in the free-energy surface spanned by the distance and the radius of gyration [Fig.~\ref{fig:cln}($c$)], where we can see only the folded, intermediate, and unfolded states and remaining are possibly degenerate.

In Fig.~\ref{fig:cln}, we can see the lower-lying free-energy basins in the reweighted stochastic embeddings are captured by both \textsc{mrse} and \textsc{stke}. We can also notice a slight difference between the metastable states lying higher in free energy. Specifically, \textsc{mrse} captures more states below a threshold of 25 kJ/mol in comparison to the embedding rendered by \textsc{stke}, in which the rest of the states is placed over 25 kJ/mol (i.e., mainly different unfolded states).

In our simulations, we do not observe a misfolded state of CLN025 shown to be highly populated in several studies~\cite{satoh2006folding,lindahl2014accelerated} employing different force fields (Amber99~\cite{wang2000well} and Amber99-SB~\cite{hornak2006comparison}, respectively) compared to CHARMM27 here~\cite{charmm}. This misfolded state is also not observed in the long unbiased simulation from Ref.~\citenum{lindorff2011fast} that employs the same CHARMM27 force field as we do.

Comparing the free-energy barriers between the different embeddings in Fig.~\ref{fig:cln}, we can see that they are similar, particularly for the \textsc{mrse} embedding and the free-energy surface spanned by the distance and the radius of gyration, i.e., from 10 to 15 kJ/mol. We can compare our results to the unbiased simulation data from the study of Lindorff-Larsen et al.~\cite{lindorff2011fast} where the authors perform a very long simulation and observe a significant number of folding and unfolding events, thus allowing unbiased estimates of free-energy barriers to be obtained. In their study, CLN025 was shown to be a ``fast folder'' with the corresponding free-energy barrier of ${\sim}10$ kJ/mol. Similar estimates have also been obtained in Ref.~\citenum{palazzesi2017conformational}. Therefore, we can conclude that the free-energy barriers in the embeddings agree well with previous computational studies.

Note that the simulation of CLN025 performed in Ref.~\citenum{lindorff2011fast} is ${\sim}100$ $\mu$s long compared to our 1-$\mu$s simulation. This clearly illustrates the great benefit of combining manifold learning with the ability to learn from biased data sets.

Overall, both the separation of the CLN025 metastable states and the free-energy landscapes calculated for the low-dimensional embeddings suggest that the proposed framework can be used to find slow CVs and physically valid free-energy estimates. The presented results (Fig.~\ref{fig:cln}) clearly show that using our approach, we can construct a meaningful and informative low-dimensional representation of a dynamical system from a biased data set, also when employing strong biasing (i.e., the high bias-factor regime in the case of well-tempered metadynamics).

We underline that diffusion reweighting makes learning CVs from high-dimensional samples possible regardless of which conformational variable is biased to generate the data set. This extends the applicability of manifold learning methods to atomistic trajectories of any type (unbiased and biased) and makes it possible to learn CVs from a biased data set where the sampling is faster and more evident than in an unbiased data set.

\section{Conclusions}
Nonlinear dimensionality reduction has been successfully applied to high-dimensional data without dynamical information. Dynamical data is a unique problem with different characteristics than generic data. Standard dimensionality reduction employed in analyzing dynamical data may result in a representation that does not contain dynamical information. This problem is even more pronounced in enhanced sampling, where we sample a biased probability distribution and additional assumptions on data structure have to be made. As such, manifold learning methods require a framework with several modifications that would allow working on trajectories obtained from enhanced sampling simulations. In this work, we introduce such a framework.

The main result of our work is deriving the reweighting procedure for manifold learning methods that use transition probabilities for building low-dimensional embeddings. These advancements enable us to directly construct a low-dimensional representation of CVs from enhanced sampling simulations. We show how our approach can be leveraged to reconstruct slow CVs from enhanced sampling simulations even in high bias-factor regimes. Our framework can be further exploited in constructing a low-dimensional representation for dynamical systems using other manifold learning methods. For instance, it could be used in spectral embedding maps~\cite{belkin2001laplacian,belkin2003laplacian} or stochastic neighbor embedding (e.g., $t$-\textsc{sne})~\cite{hinton2002stochastic,maaten2008visualizing,maaten2009learning}. There are numerous stages at which such methods have scope for different algorithmic choices. Consequently, many possible algorithms can work within our framework.

An interesting direction for further research is to combine diffusion reweighting with a metric different from Euclidean distance, for instance, by considering a metric that enables introducing a lag time, as done in the case of kinetic and commute maps~\cite{noe2015kinetic,noe2016commute,tsai2021sgoop}, a Mahalanobis kernel~\cite{singer2008non,evans2021computing}, or delay coordinates~\cite{berry2013time}. Diffusion reweighting can be extended to yield intrinsic timescales directly from enhanced sampling simulations based on their relation to eigenvalues. We plan to take this road soon.

We underline that the presented diffusion reweighting can be used in any enhanced sampling method as the method can work with any functional form of the weights. For instance, tempering methods such as parallel tempering~\cite{parallel_tempering} can be used, where the weights are given as $\e^{-\Delta\beta U}$ for the difference in the inverse temperatures $\Delta\beta$ between the simulation temperature and the target temperature.

A point that requires further addressing is the selection of variables for a high-dimensional configuration space that carry enough information about the system dynamics to characterize a low-dimensional manifold. This issue is fundamental when using the configuration variables other than the microscopic coordinates. The configuration variables do not necessarily need to be optimal. We do not have to know whether all of the chosen configuration variables are relevant for the studied process; some of them may be spurious and thermodynamically meaningless. The primary assumption in selecting such configuration variables is that some are relevant and capture slower timescales of the studied process. This assumption can be validated by using diffusion map (with reweighting if the samples are biased) to check if there is a clear separation of timescales and the dynamics of the selected configuration variables is slower compared to other variables~\cite{zwanzig1961memory,bussi2006equilibrium}. Our framework can be used for this aim; therefore, we plan to investigate the effect of selecting the configuration variables on constructing the low-dimensional CVs in the future.

Our framework makes it possible to generate biased data sets that, given the construction of enhanced sampling methods, sample a larger conformational space than standard atomistic simulations and use such data to learn low-dimensional embeddings. If a data set entails many infrequent events, the low-dimensional representation is more prone to encode them quantitatively. Moreover, in the case of the reweighted stochastic embedding methods, which we cover here, the generated embeddings can be used for biasing in an iterative manner, e.g., where we iterate between the learning and biasing phases. We believe that the accurate construction of the Markov transition probability matrix is a crucial element in implementing such an algorithm optimally without being restricted by kinetic bottlenecks (i.e., low-probability transition regions).

Overall, we expect that our approach to manifold learning from enhanced sampling simulations opens a variety of potential directions in studying metastable dynamics that can be explored.

\begin{suppinfo}
The Supporting Information is available free of charge at \url{https://pubs.acs.org/}.
\begin{itemize}
  \item Details of data sets: one-dimensional potential, alanine dipeptide, and chignolin.
  \item Additional figures: density of samples along coordinate $x$ for the one-dimensional potential; diffusion coordinates and effective timescales for alanine dipeptide.
\end{itemize}
\end{suppinfo}

\section*{Notes}
The authors declare no competing financial interest.

\section*{Acknowledgements}
J.R. acknowledges funding from the Polish Science Foundation (START), the National Science Center in Poland (Sonata 2021/43/D/ST4/00920, ``Statistical Learning of Slow Collective Variables from Atomistic Simulations''), and the Ministry of Science and Higher Education in Poland. M.C. and T.K.G. acknowledge the support of Purdue Startup Funding. Calculations were performed on the Opton cluster at the Institute of Physics, NCU.

\appendix
\section{Diffusion Reweighting}
\label{app:diffrew}
Consider a data set $\set[\big]{ \bx_k, w(\bx_k) }_{k=1}^K$ where each sample $\bx_k$ is high-dimensional and the number of samples is given by $K$ [\eq{eq:data}].

A discrete probability distribution for a stochastic process with a discrete state space is given by:
\begin{equation}
  \label{eq:dist}
  n(\bx) = \sum_l w(\bx_l) \delta\paren{\bx-\bx_l},
\end{equation}
where $\sum_k w(\bx_k) = 1$. Assuming a Gaussian kernel $G_\varepsilon$, we can account for the statistical weights to obtain the unbiased kernel density estimate [\eq{eq:rhop}]:
\begin{align}
  \label{eq:unb_density}
  \rho(\bx_k) &= \int\dx G_\varepsilon(\bx_k,\bx)n(\bx) \nonumber\\
   &= \int\dx G_\varepsilon(\bx_k,\bx)\sum_l w(\bx_l) \delta\paren{\bx-\bx_l} \nonumber\\
  &= \sum_l w(\bx_l) G_\varepsilon(\bx_k,\bx_l),
\end{align}
where the Dirac delta function $\delta\paren{\bx-\bx_l}$ leaves only the $l$-th terms from the integral. Then, up to a normalization constant, the diffusion-map kernel is given by:
\begin{equation}
  \label{eq:kappa}
  \kappa(\bx_k,\bx_l) \propto \frac{G_\varepsilon(\bx_k,\bx_l)}{[\rho(\bx_k)]^\alpha [\rho(\bx_l)]^\alpha},
\end{equation}
where the parameter $\alpha$ is called the anisotropic diffusion parameter. The normalization constant $d(\bx)$ for \eq{eq:kappa} can be calculated similarly to \eq{eq:unb_density}:
\begin{align}
  \label{eq:norm_diff}
  d(\bx_k) &= \int\dx \kappa(\bx_k,\bx)n(\bx) \\
         &= \sum_l w(\bx_l) \kappa(\bx_k,\bx_l).
\end{align}
A Markov operator $\mathcal{M}$ acting on an auxiliary function $f(\bx)$ can be written as:
\begin{equation}
  \label{eq:oper}
  \paren[\big]{\mathcal{M}f}(\bx_k)=\int\dx \mathcal{K}(\bx_k,\bx)f(\bx)n(\bx),
\end{equation}
where $\mathcal{K}$ is known as a kernel of the Markov operator $\mathcal{M}$ and $\int\dx \mathcal{K}(\bx_k,\bx)n(\bx)=1$. Using the above definition, we can evaluate the Markov transition matrix $M(\bx_k,\bx_l)$ by acting the Markov operator $\mathcal{M}$ on the function $f(\bx_k)$ using the anisotropic diffusion kernel [\eq{eq:kappa}] as $\mathcal{K}$ in \eq{eq:oper}:
\begin{align}
  \label{eq:mmm}
  \paren[\big]{\mathcal{M}f}(\bx_k) &= \int\dx \frac{\kappa(\bx_k,\bx)}{d(\bx_k)} f(\bx_k)n(\bx) \nonumber\\
   &= \int\dx\frac{\kappa(\bx_k,\bx)}{d(\bx_k)} f(\bx_k) \sum_l w(\bx_l)\delta\paren*{\bx-\bx_l} \nonumber\\
  &= \sum_l \frac{w(\bx_l)\kappa(\bx_k,\bx_l)}{\sum_n w(\bx_n) \kappa(\bx_k,\bx_n)} f(\bx_k) \nonumber\\
  &= \sum_l M(\bx_k,\bx_l)f(\bx_k),
\end{align}
which gives us the definition of the Markov transition matrix $M$.

By introducing a rescaled statistical weight:
\begin{equation}
  w_\alpha(\bx)=\frac{w(\bx)}{\bkt*{\rho(\bx)}^\alpha},
\end{equation}
we can write $M(\bx_k,\bx_l)$ as:
\begin{align}
  \label{eq:markov2}
  M(\bx_k,\bx_l) = \frac{w_\alpha(\bx_l)G_\varepsilon(\bx_k,\bx_l)}{\sum_n w_\alpha(\bx_n)G_\varepsilon(\bx_k,\bx_n)}.
\end{align}
Therefore, a general expression for the reweighting factor can be given as:
\begin{equation}
  \label{eq:rew}
  r(\bx_k,\bx_l) = w_\alpha(\bx_k)w_\alpha(\bx_l),
\end{equation}
where $w_\alpha(\bx_k)$ is canceled out in \eq{eq:markov2} during the normalization. Alternatively, we can express \eq{eq:rew} using a biased density estimate $\rho_V(\bx_k)=\sum_l G_\varepsilon(\bx_k,\bx_l)$ [\eq{eq:rhobiased}]:
\begin{align}
  \label{eq:approx}
  \rho(\bx_k) &= \sum_l w(\bx_l) G_\varepsilon(\bx_k,\bx_l) \nonumber\\
         &\approx w(\bx_k) \sum_l  G_\varepsilon(\bx_k,\bx_l) \nonumber\\
         &= w(\bx_k) \rho_V(\bx_k),
\end{align}
which is similar to the standard reweighting formula. Using \eq{eq:approx} and setting $\alpha=\frac{1}{2}$, we obtain:
\begin{equation}
  \label{eq:rrr}
  r(\bx_k,\bx_l) = \sqrt{\frac{w(\bx_k)}{{\rho_V(\bx_k)}}} \sqrt{\frac{w(\bx_l)}{{\rho_V(\bx_l)}}},
\end{equation}
which concludes the derivation of \eq{eq:rfactorslow}.

\section{Square-Root Approximation}
\label{app:sqra}
Here, we want to derive \eq{eq:entropy} by considering approximations to the Markov transition matrix used in \textsc{stke}~\cite{zhang2018unfolding}.

As we discuss in Sec.~\ref{sec:rse}, we want to obtain a format of the transition matrix similar to that of the square-root approximation to the Fokker-Planck operator. We start from the reweighting factor given by \eq{eq:stke} and construct the following Markov transition matrix:
\begin{equation}
  \label{eq:mmstke}
  M(\bx_k,\bx_l) = \frac{{\sqrt{\rho(\bx_l)}} G_\varepsilon(\bx_k,\bx_l)}{\sum_n \sqrt{\rho(\bx_n)} G_\varepsilon(\bx_k,\bx_n)},
\end{equation}
where $\rho(\bx_k)$ is canceled out due to the normalization. By assuming that $\varepsilon$ is sufficiently small, we can take the following approximation to the normalization constant of \eq{eq:mstke}:
\begin{equation}
  \label{eq:normconst}
  \mathcal{Z_\rho}=\sum_n \sqrt{\rho(\bx_n)} G_\varepsilon(\bx_k,\bx_n) \propto \sqrt{\rho(\bx_k)},
\end{equation}
where we approximate the average of local densities under the kernel density estimate by the density centered on $\bx_k$. Then, \eq{eq:mstke} is:
\begin{equation}
  M(\bx_k,\bx_l) = \sqrt{\frac{\rho(\bx_l)}{\rho(\bx_k)}} G_\varepsilon(\bx_k,\bx_l),
\end{equation}
which gives us a relation similar to the square-root approximation of the infinitesimal generator of the Fokker-Planck operator~\cite{lie2013square,heida2018convergences,donati2018estimation,kieninger2020dynamical} [\eq{eq:entropy}].

\section{Gaussian Mixture for the Markov Transition Matrix $M$}
\label{app:gaussmix}
Here, we describe a procedure used to automatically estimate bandwidths for a Gaussian mixture used in \textsc{mrse}. The procedure is similar to that used in $t$-\textsc{sne}, with the exemption of using a Gaussian mixture instead of a single Gaussian and expanding the procedure to account for the statistical weights. We follow a procedure outlined in Ref.~\citenum{rydzewski2020multiscale}.

We use a Gaussian mixture to represent the Markov transition matrix [\eq{eq:mmrse}]. Each Gaussian has a positive parameter set $\boldsymbol{\varepsilon}=\set{\varepsilon_k}$. We find the appropriate values of $\boldsymbol{\varepsilon}$ so that the Shannon--Gibbs entropy of each row of $M(\bx_k,\bx_l)\equiv(p_{kl})$, $s_k = -\sum_l p_{kl} \log p_{kl}$ is approximately equal to the number of neighbors $n_p$ given as the logarithm of perplexity~\cite{maaten2008visualizing}.

Considering the weights of the exponential form, $w_k=\e^{\beta V_k}$, where $V_k$ is the relative bias potential at the $k$-th sample [\eq{eq:coft}], the entropy for the $k$-th row of the Markov transition matrix $M$ has has to be corrected by including the bias potential in comparison to that used in $t$-\textsc{sne}~\cite{maaten2008visualizing}. The bias-free term is given by:
\begin{equation}
  \label{eq:entr}
  s_k = \log \sum_l p_{kl} + \varepsilon_k \sum_l p_{kl} \enorm{\bx_k-\bx_l},
\end{equation}
and the correction term is:
\begin{equation}
  \label{eq:entr}
  s'_k = -\frac{\beta}{2}\paren*{\sum_l p_{kl} V_l + V_k},
\end{equation}
where the sum is the averaged bias potential with respect to the transition probabilities of the Markov transition matrix $M$.

Therefore, the optimization of $\boldsymbol{\varepsilon}$ is performed by finding such $\set{\varepsilon_k}$ so it minimizes the difference between the Shannon--Gibbs entropy for $k$-th row of $M$ and the number of neighbor in a manifold:
\begin{equation}
  \varepsilon_k = \operatorname{min}_{\varepsilon} (s_k +s'_k - n_p)
\end{equation}
which can be solved using binary search. After finding the set $\boldsymbol{\varepsilon}$ of bandwidths (each for a single row of $M$) for a perplexity value, we can calculate the Gaussian mixture representation of $M$ as an average over $M$ estimated for each selected perplexity. Perplexities for each $M$ matrix can be also estimated automatically.

A detailed derivation and a discussion about the procedure outlined here, can be found in Ref.~\citenum{rydzewski2020multiscale}.

\section{Landmark Sampling: Selecting Training Set}
\label{app:wtrandom}
For \textsc{stke}, the training data set is selected using a geometric subsampling scheme that results in landmarks distributed uniformly. Specifically, the training data set is created such that $\min_{kl} \| \bx_k - \bx_l \| \ge r_c$, where $r_c$ is a minimal pairwise distance, which modifies the level of sparsity for building the Markov transition matrix.

In \textsc{mrse}, we use weight-tempered random sampling in which the training data set is selected according to statistical weights. The statistical weights are scaled, $w^{1/\tau}$, where $\tau\ge 1$ is the tempering parameter, and samples are selected according to the scaled weights. It has been shown that in the limit of $\tau\rightarrow\infty$, we obtain the biased marginal probability, and for $\tau\rightarrow 1$, we recover the unbiased probability. A detailed discussion with a comparison to other landmark sampling algorithms is provided in Ref.~\citenum{rydzewski2020multiscale}.

\section{Parameters for Reweighted Stochastic Embedding}
\label{app:comparison}
\setlength{\tabcolsep}{10pt}
We show a summary of the reweighted stochastic embedding methods and parameters in Tab.~\ref{tab:comp}. Note that many parameters for the reweighted stochastic embedding methods are set as in Refs.~\citenum{rydzewski2020multiscale,zhang2018unfolding}.

\begin{table*}[t]
  \caption{Summary of reweighted stochastic embedding methods used in Sec.~\ref{sec:rse}: multiscale reweighted stochastic embedding (\textsc{mrse}) and stochastic kinetic embedding (\textsc{stke}). Labels: reweighting factor $r(\bx_k,\bx_l)$, high-dimensional Markov transition matrix $M(\bx_k,\bx_l)$, low-dimensional Markov transition matrix $Q(\bz_k,\bz_l)$.}
  \begin{tabular}{p{0.22\textwidth}p{0.35\textwidth}p{0.31\textwidth}}
    \hline
    & \textsc{mrse} & \textsc{stke} \\
    \hline\hline

    Reweighting factor $r(\bx_k,\bx_l)$
      & $\sqrt{w(\bx_k) w(\bx_l)}$ [\eq{eq:rmrse}]
      & $\sqrt{\rho(\bx_k)\rho(\bx_l)}$ [\eq{eq:stke}] \\

    High-dim. prob. $M(\bx_k,\bx_l)$
      & Gaussian mixture [\eq{eq:mmrse}] for perplexities $\in\set{256,128,64,32}$
      & Gaussian [\eq{eq:mstke}] with $\varepsilon=0.12$ \\

    Low-dim. prob. $Q(\bz_k,\bz_l)$
      & $t$-distribution [\eq{eq:tdist}]
      & Gaussian with $\varepsilon=0.12$ \\

    Landmark sampling
      & Weight-tempered random sampling for $\tau=3$ and 5000 landmarks
      & Minimal pairwise distance for $r_c=1.2$ and 97000 landmarks\\

    Activation functions
      & Hyperbolic tangent (3 layers)
      & ReLU (3 layers) \\

    Optimizer
      & Adam $(\mu=0.001, \beta_1=0.9, \beta_2=0.999)$
      & Adam $(\mu=0.001, \beta_1=0.9, \beta_2=0.999)$ \\

    Batch size
      & 1000
      & 256 \\

    \hline
  \end{tabular}
  \label{tab:comp}
\end{table*}

\bibliography{main}

\end{document}